\documentclass[preprint,5p]{elsarticle}

 \usepackage{epsfig}

\usepackage{amssymb}

\usepackage{color}

\usepackage{lineno}

\biboptions{square,sort&compress}

\journal{Nuclear Physics A}
\def\NIM{\em Nucl.\ Instr.\ and Meth.}
\def\NIMA{{\em Nucl.\ Instr.\ and Meth.}\ A}
\def\JCAP{\em J.\ Cosmol.\ Astropart.\ Phys.}
\def\Journal#1#2#3#4{{#1} {\bf #2}, #3 (#4)}

\begin{document}

\begin{frontmatter}



\title{Studies of Acoustic Neutrino Detection Methods with ANTARES}


\author[ECAP]{K.~Graf}\ead{kay.graf@physik.uni-erlangen.de}\author{\\on
  behalf of the ANTARES Collaboration} \address[ECAP]{ECAP, University
  of Erlangen-Nuremberg, Erwin-Rommel-Str.\ 1, D-91058 Erlangen,
  Germany}
\begin{abstract}
  The emission of neutrinos within a wide energy range is predicted
  from very-high-energy phenomena in the Universe.  Even the current
  or next-generation Cherenkov neutrino telescopes might be too small
  to detect the faint fluxes expected for cosmic neutrinos with
  energies exceeding the EeV scale. The acoustic detection method is a
  promising option to enlarge the discovery potential in this
  highest-energy regime.  In a possible future deep-sea detector, the
  pressure waves produced in a neutrino interaction could be detected
  by a $\gtrsim 100\,$km$^3$-sized array of acoustic sensors, even if
  it is sparsely instrumented with about 100\,sensors/km$^3$.  This
  article focuses on the AMADEUS set-up of acoustic sensors, which is
  an integral part of the ANTARES detector. The main aim of the
  project is a feasibility study towards a future acoustic neutrino
  detector. However, the experience gained with the ANTARES-AMADEUS
  hybrid opto-acoustic set-up can also be transferred to future very
  large volume optical neutrino telescopes, especially for the
  position calibration of the detector structures using acoustic
  sensors.
\end{abstract}

\end{frontmatter}
\section{Introduction}
At ultra-high energies (UHE, $E \gtrsim 10^{18}\,$eV), neutrinos are
the only viable messengers beyond the ``local'' universe.  Photons
interact with the microwave and infrared background light. Protons
suffer energy losses due to photo-production of pions (GZK mechanism
\cite{grei,zats}) and of e$^+$e$^-$ pairs; nuclei are additionally
subject to photo-nuclear reactions. All these reactions confine
undisturbed propagation of UHE cosmic rays to distances below 100\,Mpc
\cite{allard}.

UHE neutrinos, on the other hand, are undisturbed and ``guaranteed''
by the GZK mechanism. The interaction rate of neutrinos originating
from the propagation of UHE protons in 1\,km$^3$ water equivalent has
been estimated to be about 0.2 per year \cite{seckel}.  Consequently,
target masses exceeding 100\,km$^3$ are required to obtain a few
neutrino events per year. Such detector sizes are not reachable with
current technologies and new detection techniques have to be
considered for the study of UHE cosmic neutrinos, e.g.\ the acoustic
method discussed here.

The investigation of acoustic neutrino detection has historically
evolved in the context of Cherenkov neutrino {te\-le\-scopes}. The
method was discussed in the 1970s within the DUMAND project
\cite{bowen}. Acoustic set-ups have been integrated in the framework
of existing Cherenkov experiments in the 2000s: AMADEUS in
{ANTA\-RES}, ONDE in NEMO, SPATS in IceCube and acoustic test set-ups
in the lake Baikal neutrino telescope\footnote{There is a variety of
  additional ongoing projects - not only in sea water but also in ice
  and salt, cf.\ \cite{arena} for an overview.}. A next step towards
an acoustic neutrino detector could be the integration of acoustic
sensors into a next-generation neutrino telescope like KM3NeT
\cite{km3net}.
\section{Acoustic Neutrino Detection}\label{sec:acou_detection}

The acoustic detection method is based on the reconstruction of
characteristic sound pulses that are generated by neutrino-induced
particle cascades in water. The generation mechanism is described by
the \emph{thermo-acoustic model} \cite{ask2,lear}, which connects the
energy deposition by the cascade particles to a local heating
accompanied by an expansion of the medium.  This translates into a
pressure wave which propagates in a wave pattern of cylindrical
symmetry through the medium, expanding in time perpendicular to the
cascade axis.  The thermo-acoustic signal is bipolar in time with a
peak-to-peak amplitude of the order of 10\,mPa per 1\,EeV cascade
energy at a vertical distance of 200\,m from the cascade \cite{aco2}.
The peak-to-peak signal length is several tens of microseconds,
resulting in a spectral energy density peaked at about 10\,kHz.

The {\it attenuation length} of the sonic wave in sea water is of the
order of 5\,km (1\,km) at 10\,kHz (20\,kHz), much longer than the one
of the Cherenkov light used in optical detectors.  A sensor density of
the order of 100\,sensors/km$^3$ is sufficient for event
reconstruction in an acoustic detector. This density is driven by the
geometry of the sound emission and not by the attenuation length. The
simulated effective volume for an array of 200 acoustic
storeys\footnote{A signal detection threshold of 5\,mPa is assumed in
  this study.}  (cf.\ Sec.\ \ref{sec:amadeus}), randomly distributed
in 1\,km$^3$, reaches 1, 10 and 100\,km$^3$ at 20, 60 and 600\,EeV,
respectively \cite{karg}.  The effective volume for cascades of the
1\,km$^3$-sized IceCube neutrino telescope with roughly 5000 optical
sensors is $\approx 3$\,km$^3$ at 1\,EeV and rises by $\approx
0.15\,$km$^3$ per decade in energy \cite{halzen}. A direct comparison
of these numbers is not fully justified as the first simulation is
without quality cuts on the reconstruction of the acoustic wave and
for a detector size that is comparable to the sound range and
therefore unfavourable for acoustic detection. The numbers can,
however, be regarded as a guideline.  The break-even point is at about
50\,EeV, so below the EeV-scale an optical detector is more sensitive.
At the EeV-scale a hybrid opto-acoustic detector could combine the
advantages of both detection methods; in ice the radio technique could
add to the sensitivity \cite{vand}. However, to measure neutrinos in
water with energies beyond the EeV-scale an extremely large acoustic
detector would be necessary.

In a hypothetical $\gtrsim 100\,$km$^3$ acoustic detector, the
signature of a neutrino-induced sound pulse has to be recognised in
the abundant ambient acoustic noise; the noise level of the order of
10\,mPa (at calm sea) in the frequency range of the acoustic pulse
sets the energy threshold for acoustic detection in the EeV range. In
addition to the persistent noise, neutrino-generated signals have to
be distinguished from transient background pulses which can originate
from either surface or under-sea sound sources (fauna or
anthropogenic).  The AMADEUS project has been launched for the
investigation of these points.

\section{The AMADEUS project}
\label{sec:amadeus}
The main goal of the AMADEUS ({\bf A}ntares {\bf M}odules for {\bf
A}coustic {\bf DE}tection {\bf U}nder the {\bf S}ea)
project\,\cite{lahm} is to conclude on the feasibility of acoustic UHE
neutrino detection in large, sea-based acoustic detector arrays.  In
addition, the integration of AMADEUS in the ANTARES neutrino telescope
\cite{ant1,antres} allows for studying hybrid opto-acoustic detection
techniques.

\begin{figure}
 \centering
 \psfig{figure=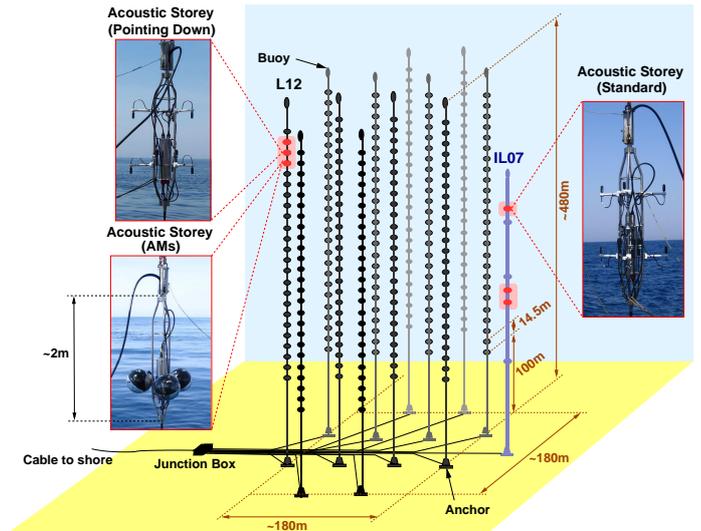,width=9cm}
 \caption{Schematic view of the ANTARES detector \cite{lahm}.  The six
   Acoustic Storeys are highlighted and their three different set-ups,
   implemented to test acoustic detector designs, are shown.}
 \label{fig:antares_schematic}
\end{figure}

The ANTARES experiment (Fig.~\ref{fig:antares_schematic}) is located
off-shore at a water depth of about 2500\,m in the Mediterranean Sea,
about 40\,km south-east of Toulon (France). The detector comprises 12
detection lines labelled {\em L1 \--- L12}\,; an additional line ({\em
  IL07}\,) is instrumented with devices for environmental monitoring.
A detection line holds 25 {\em storeys}, each housing three PMTs
\cite{ant_om,ant_pmt}, auxiliary sensors and read-out hardware.  Each
line is fixed to the sea floor by an anchor and kept vertically by an
undersea buoy.
The acoustic set-up is integrated into the detector as a set of six
{\em Acoustic Storeys} ({\em ASs}\,), which are modified versions of
standard ANTARES storeys, at depths from 2050 to 2300m. In an AS, the
three PMTs are substituted by six acoustic sensors and custom-designed
electronics is used for the digitisation and pre-processing of the
analogue signals. Once digitised at the storey level, the acoustic
data are fed into the detector data stream and are further processed
by the data-acquisition hard- and software of ANTARES \cite{ant_daq}.
On-line filters using acoustic signal processing algorithms like
matched filters \cite{turin} select and store events of interest on a
dedicated server cluster on-shore.  These filters reduce the data
volume by a factor of more than 100, from 1.5\,TByte per day arriving
from the ASs to about 10\,GByte per day stored on hard-disk for
off-line analysis. Both filter and analysis algorithms are
implemented in standard ANTARES software.

The three acoustic storeys on the IL07 started operation in December
2007, those on L12 with the completion of ANTARES in May 2008.
AMADEUS is now fully functional with 34 active sensors.  After two
years of operation, the system has demonstrated excellent long-term
stability and data-taking characteristics. During the ANTARES
data-taking periods, the AMADEUS set-up has been continuously active
for about 85$\%$ of the time. Studies of ambient noise features, of
the calibration of the set-up and of source position reconstruction
have been conducted \cite{arena}. Current studies cover e.g.\ temporal
and spatial source distributions, signal classification and mammal
tracking.

\section{Acoustic Positioning and \\Opto-Acoustic Modules}

The positioning of the sensors of an under-sea neutrino detector is
essential, as its structures cannot be rigid and thus move in the
deep-sea currents.  To achieve best pointing accuracy, a precision of
the order of 1\,cm is needed for acoustic detection and of 10\,cm for
optical systems.  The positioning in ANTARES is performed via acoustic
triangulation using acoustic emitters at the bottom of each line and
acoustic sensors\footnote{Specialised sensors not suitable for
  acoustic detection.} at every fifth storey, supported by compass and
tilt-meter data recorded at each storey.

\begin{figure}
 \centering
 \psfig{figure=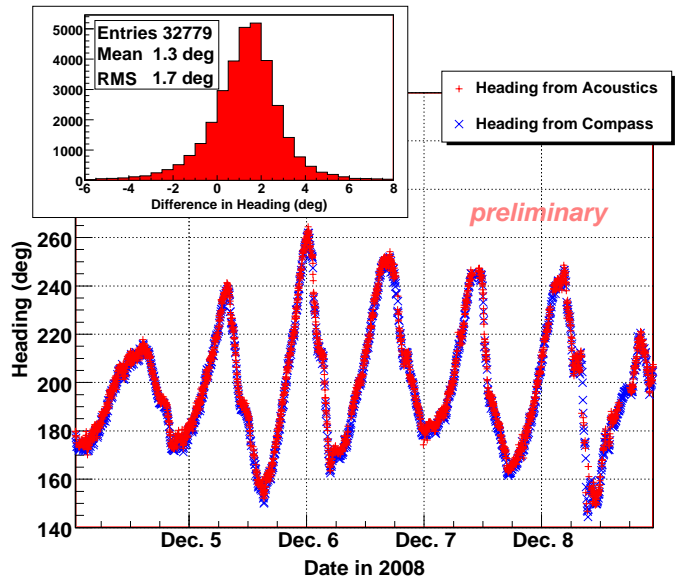,width=9.5cm}
 \caption{Heading (northing) of an acoustic storey (top-most on IL07)
   from compass data (labelled {\color{blue}$\times$}) and from the
   positioning of the acoustic sensors (labelled
   {\tiny{\color{red}$+$}}).  The oscillating pattern in these five
   days is due to well-known Coriolis-induced current variations.}
 \label{fig:heading}
\end{figure}

The ability to position the acoustic storeys is shown in
Fig.~\ref{fig:heading}, where the heading measured by the compass in
one storey is compared to the heading derived from the individually
reconstructed position of the acoustic sensors on that storey. The RMS
of the deviation (1.7$^\circ$) can be converted to a positioning
uncertainty of a few centimetres.

Based on the positive results gained in the operation of one storey
equipped with {\it acoustic modules} ({\it AMs}\,)\footnote{An AM
  consists of two piezo-ceramic sensors with preamplifiers, glued to
  the inside of a glass sphere usually housing a PMT.}, the idea of
combined {\it opto-acoustic modules} ({\it OAMs}), housing both PMTs
and acoustic sensors, has evolved.  The acoustic sensors in OAMs can
be used for various purposes: for the positioning of the optical
sensors in a future optical neutrino telescope, but also for acoustic
detection and marine science (e.g.\ mammal tracking \cite{nemo}).

First laboratory tests of a parallel operation of a PMT with its
high-voltage supply and acoustic sensors within one glass sphere have
shown no significant degradation of performance of either of the two
devices.  The advantages of OAMs, as compared to separate optical and
acoustic sensors, are a simplified data-acquisition hardware, the use
of the optical data handling also for the acoustic data (as
demonstrated with AMADEUS) and no need for components outside the
sphere, thus avoiding potentially problematic penetrators through the
glass and additional containers.  However, the signal quality when
coupling acoustic sensors to the inside of a glass sphere is reduced
compared to dedicated acoustic sensors. First-principle considerations
show that in combined OAMs the orientation could be reconstructed with
uncertainties of 10$^{\circ}$ when using two acoustic sensors per OAM
at a distance of 25\,cm. The resulting precision for positioning is in
agreement with the requirements for an optical detector.

\section{Conclusions}
Cosmic neutrinos are the only viable UHE messengers beyond the
``local'' universe.  The acoustic neutrino detection technique is an
option to extend neutrino astrophysics to the extreme energy range,
either in stand-alone detectors or in combined opto-acoustic sensor
arrays.  The AMADEUS system, dedicated to the investigation of this
technique, has been successfully integrated into the ANTA\-RES
neutrino telescope.  Except for size, the system has all features
required for an acoustic detector and thus allows for deciding on the
feasibility of neutrino detection with a potential future large
acoustic sensor array. In the context of AMADEUS it is currently being
investigated whether an integrated opto-acoustic sensor configuration
has the potential to combine the position calibration of an optical
detector with acoustic detection studies and marine science
applications.

\section*{Acknowledgements}
The presented work is performed within the ANTARES collaboration and
the author is supported by the German Ministry for Education and
Research (BMBF) by Grants 05CN5WE1/7 and 05A08WE1. The author wishes
to compliment the organisers on a most interesting workshop.

\section*{References}
\bibliographystyle{elsarticle-num}

\end{document}